# Isotropic Band Gaps and Freeform Waveguides Observed in Hyperuniform Disordered Photonic Solids

Weining Man[1*], Marian Florescu[2], Eric Paul Williamson[1], Yingquan He[1], Seyed Reza Hashemizad[1], Brian Y.C. Leung[1], Devin Robert Liner[1], Salvatore Torquato[3,4,5], Paul M. Chaikin[6], and Paul J. Steinhardt[3,4*]

[1] Department of Physics and Astronomy, San Francisco State University, San Francisco, CA 94132, USA

[2] Advanced Technology Institute and Department of Physics, University of Surrey, Guildford, Surrey GU2 7XH, United Kingdom

[3] Department of Physics, Princeton University, Princeton, NJ 08544, USA

[4] Princeton Center for Theoretical Science, Princeton University, Princeton, NJ 08544, USA

[5] Department of Chemistry, Princeton University, Princeton, NJ 08544, USA

[6] Department of Physics, New York University, New York, NY 20012, USA

*Correspondence to: weining@sfsu.edu; steinh@princeton.edu

**Abstract:** Recently, disordered photonic media and random textured surfaces have attracted increasing attention as strong light diffusers with broadband and wide-angle properties. We report the first experimental realization of an isotropic complete photonic band gap (PBG) in a two-dimensional (2D) disordered dielectric structure. This structure is designed by a constrained-optimization method, which combines advantages of both isotropy due to disorder and controlled scattering properties due to low density fluctuations (hyperuniformity) and uniform local topology. Our experiments use a modular design composed of $Al_2O_3$ walls and cylinders arranged in a hyperuniform disordered network. We observe a complete PBG in the microwave region, in good agreement with theoretical simulations, and show that the intrinsic isotropy of this novel class of PBG materials enables remarkable design freedom, including the realization of waveguides with arbitrary bending angles impossible in photonic crystals. This first experimental verification of a complete PBG and realization of functional defects in this new class of materials demonstrates their potential as building blocks for precise manipulation of photons in planar optical micro-circuits and has implications for disordered acoustic and electronic bandgap materials.

**Introduction**

The first examples of synthetic materials with complete photonic band gaps (PBGs) (1,2) were photonic crystals utilizing Bragg interference to block light over a finite range of frequencies. Due to their crystallinity, the PBGs are highly anisotropic, a potential drawback for many applications. The idea that a complete PBG (blocking all directions and all polarizations) can exist in isotropic disordered systems is striking, since it contradicts the longstanding intuition that periodic translational order is necessary to form PBGs. The paradigm for PBG formation is Bloch's theorem (3): a periodic modulation of the dielectric constant mixes degenerate waves propagating in opposite directions and leads to standing waves with high electric field intensity in the low dielectric region for states just above the gap and in the high dielectric region for states just below the gap. Long-range periodic order, as evidenced by Bragg peaks, is necessary for this picture to hold. The intrinsic anisotropy associated with periodicity can greatly limit the scope of PBG applications and places a major constraint on device design. For example, even though three-dimensional (3D) photonic crystals with complete PBGs have been fabricated for two decades (4), 3D waveguiding continues to be a challenge. Very recently, Noda's group reported the first successful demonstration of 3D waveguiding (5). However, it was found that, due to the mismatch of the propagation modes in line defects along various symmetry orientations, vertical-trending waveguides have to follow one particular major symmetry direction to effectively guide waves out of the horizontal symmetry plane in a 3D woodpile photonic crystal (5).

Recently, disordered photonic media and random textured surfaces have attracted increasing attention as strong light diffusers with broadband and wide-angle properties (6--9). Disorder is conventionally thought to wipe out energy band gaps and produce localization and diffusive transport, an exciting research area of its own (10-15). Although there are examples of disordered electronic systems with large band gaps, most notably amorphous silicon, complete photonic band gaps are more difficult to achieve due to the polarization differences. It is especially difficult for 2D structures to have energy gaps in both polarizations that overlap. In 2D structures, the two polarizations, with the electric field parallel (transverse electric, TE) or perpendicular (transverse magnetic, TM) to the 2D plane, behave completely differently depending on whether the E field is parallel to dielectric boundaries or not (3,16). In 3D, there is no mirror symmetry to allow TM/TE separation, and in common 3D PBG structures (i.e., woodpile, diamond-like, inverse opal) the effective dielectric distribution seen by different polarizations of light propagating in the same direction is rather similar. Notomi et al. (17,18) have discussed 3D photonic amorphous diamond structures that appear to have PBGs based on studies of small samples, although systematic convergence tests using samples of increasing size confirming that complete PBGs persist have not yet been performed. However, for 2D structures, the perfect long-range and short-range order in various 2D photonic crystals is often not sufficient for forming a complete PBG structures even at dielectric contrast ratio as high as 11.5(Si *vs.* air) (3). The best-known exception is a triangular lattice of large air holes in Si (3).

Nevertheless, Florescu et al. (19) have recently devised an algorithm for constructing disordered 2D arrangements of dielectric materials with substantial band gaps, comparable to those in the best photonic crystals at the same dielectric contrast (19). 2D photonic solids with complete PBGs are of practical significance, since most microcircuit designs are based on planar architectures (20,21). This structure is designed by a novel constrained-optimization method, which combines advantages of both isotropy due to disorder and controlled scattering properties due to low density fluctuations (hyperuniformity) and uniform local topology (19). The key features of the design are: (1) a disordered network of dielectric cylinders and walls in which each cylinder is connected to three neighbors (trivalency); and (2) an arrangement of the cylinder centers in a hyperuniform point pattern, where the number variance of points in a "window" of radius $R$, $\sigma(R) = \langle N_R^2 \rangle - \langle N_R \rangle^2$, is proportional to $R$, where $N_R$ is the number of points inside the window. Note that, for a 2D random Poisson distribution, $\sigma(R) \propto R^2$, is proportional to the window area, whereas hyperuniform structures, including crystals and quasicrystals, have $\sigma(R) \propto R$. Because of these two features, the photonic design pattern has uniform nearest-neighbor connectivity and hyperuniform long-range density fluctuations (or, equivalently, a structure factor with the property $S(k) \to 0$ for wavenumber $k \to 0$) (22) similar to crystals; at the same time, the pattern exhibits random positional order, isotropy, and a circularly symmetric diffuse structure factor $S(k)$ similar to a glass.

## Results
### Demonstration of isotropic bandgap formation
Our study focuses on a subclass of 2D hyperuniform patterns with the largest band gaps for a given dielectric contrast (19); these designs, referred to as "stealthy" (23), have a structure factor $S(k)$ precisely equal to zero for a finite range of wavenumbers $k < k_C$ for some positive $k_C$. We have constructed the first physical realization of a hyperuniform stealthy design (Fig. 1) using commercially available $Al_2O_3$ cylinders and walls cut to the designed heights and widths.

For the bandgap measurements, the transmission is defined as the ratio between transmitted intensity with and without the sample in place. We first used the hyperuniform disordered structure shown in Fig. 1b, and plotted the measured transmission normal to its boundary as the blue curves in Fig. 2a (TE) and Fig. 2c (TM). Next, to check the angular dependence of the photonic properties, cylinders and walls were removed from the corners of the samples to construct a nearly circular boundary of diameter 13*a*. The samples were rotated along the axis perpendicular to the patterned plane, and the transmission was recorded every two degrees from 0 to 180 degree for both TE and TM polarizations. The average transmission over all incident angles is plotted as the red curves in Fig. 2a (TE) and Fig. 2c (TM). The regions of low transmission (20 dB relative drop compared to the measured maximum band pass transmission) agree well with the calculated TE and TM bandgaps (see below). The calculated upper boundary of the TM bandgap and lower boundary of the TE bandgap, defining the complete PBG region, are indicated with vertical dash-dot lines.

In Fig. 3, we use color contour plots to present the measured transmission, *T*, as a function of frequency and incident angle. Between the calculated boundaries (white lines) of the complete PBG, the measured transmission through the hyperuniform structure for TE (Fig. 3a) and TM (Fig. 3b) polarizations show an isotropic complete PBG (horizontal blue stripes), with a relative gap contrast deeper than -20dB. A similar square lattice constructed with the same $Al_2O_3$ cylinders and $Al_2O_3$ walls of the same thickness is measured for comparison. As expected, in the square-lattice photonic crystal, stop gaps due to Bragg scattering occur along the Brillouin zone boundaries, are anisotropic and change frequency in different directions. For TM polarization (Fig. 3d) the stop gaps in different directions are wide enough to overlap and form a PBG, while there is no band gap for TE polarization (Fig. 3c). As a further comparison, our direct band simulation shows that the champion photonic crystal structure (a triangular lattice of air holes in dielectric), with the same dielectric constant contrast of 8.76 : 1 and filling fraction of 27%, has a complete gap of 5.2%, slightly larger than the 4.1% complete gap found in our disordered structure. The triangular structure maintains the anisotropy characteristic of periodic structures: the central frequency and the width of the stop gaps along different directions vary by 24% and 44%, respectively. In contrast, for the hyperuniform disordered structure, the central frequency and the width of the stop gaps in different directions are statistically identical. The measured transmitted power at any frequency is much lower for TM polarization than for TE polarization, in both our hyperuniform sample and our square-lattice sample. For each polarization, the transmitted power is limited by the horn geometry, namely the rectangular shape, asymmetric radiation pattern, and the relatively small radiation acceptance angle of 15°. Nevertheless, for both polarizations we observe the aforementioned 20dB reduction of transmission, confirming the existence of the PBG.

Our experimental results are compared with theoretical band structure calculations obtained using a supercell approximation and the conventional plane-wave expansion method (24). The size of the supercell used in the simulations is $22a \times 22a$ (the entire region of Fig.1a). The calculated DOS (green curves in Fig 2b & 2d) for both TE and TM modes is zero within the PBG. Finite-difference time-domain (FDTD) simulations of the transmission spectrum through a finite sample of $22a \times 22a$ (blue curves in Figs. 2b & 2d) show regions of considerably reduced transmission in the spectral region of the PBGs and overlap our experimental results. Due to background dark noise (around -40dB), and the finite size of $13a \times 13a$, the experiment is limited to detecting a gap contrast of no more than 30 dB, though the simulations of the finite sample indicate suppression by six orders of magnitude.

**Demonstration of the effective freeform waveguiding**
In order to test whether light can be guided through our hyperuniform disordered structure, a straight channel was created by removing cylinders and walls within a straight strip of width $2a$, as shown in Fig. 4a. The horn antennas were placed directly against the ends of the channel for the transmission measurement. The TM transmission spectrum for the open channel is shown in Fig. 4b. The calculated TM-polarization gap is highlighted with shading. Our measurements clearly demonstrate that a broad band of frequencies is guided through the open channel. The transmission values presented in Fig. 4 are simply the ratio of the detected power over source power, with no normalization or correction for coupling loss. Considering the substantial coupling loss expected between the microwave horn antennas and the waveguide channels, the measured transmission of 20% ~25% is impressively high, much higher than values which have been routinely reported in successful wave-guiding demonstrations in photonic crystals, i. e. less than $10^{-1}$ in Ref. (25). Since the substantial coupling loss between the microwave horns and the channel openings is unknown, to evaluate the waveguiding efficiency, we carried out an experimental comparison with a straight channel of width $2a$ and the same length, which is created by removing one row of cylinders and their connecting walls in the square-lattice photonic crystal. The square-lattice photonic crystal has a wide TM-polarization PBG, and a straight line-defect in it is supposed to offer 100% transmission in the absence of coupling loss and absorption. Similar square-lattice photonic crystals have been used as standards for TM-polarization waveguiding demonstrations (25,26). We find that, under the same coupling condition, the measured transmitted energies through straight waveguides in the hyperuniform disordered structure (Fig. 4b) and the square-lattice (Fig. 4f) are quite comparable, suggesting that there is little loss of the guided mode over this length scale despite the disorder in our structures.

In photonic crystals, efficient waveguides are limited to certain directions by crystal symmetries. The disorder and isotropy of hyperuniform structures should relax many of the restrictions found in periodic structures (27). The flexibility of our experimental design makes it easy to form channels with arbitrary bending angles and to decorate their sides, corners and centers with cylinders and walls for tuning and optimizing the transmission bands. Fig. 4d shows a waveguide with a sharp 50° bend made by removing cylinders and walls within a strip of width 2*a*, keeping the boundaries and corner of the path relatively smooth. Fig. 4e shows the measured transmission, which is approximately the same as that of the straight waveguide, despite the sharp bend. An equally good result is obtained with the "S" shaped freeform waveguide shown in Fig. 4g. As with the previous channels, the transmitting and receiving horn openings are parallel to the input and output sections of the channel, respectively. Conservation of photon momentum due to translation invariance is absent in any non-straight waveguide in either periodic or disordered structures. Tuning with defects is often required to obtain effective coupling along the bending path in photonic crystals. Similarly, in our isotropic disordered structures, back scattering of the propagating mode can be alleviated by optimizing the spacing and cylinder sizes along the channel. As shown in Fig. 4h, we found that, for channels with length of tens of *a*, the transmission through such a freeform S-shape channel can achieve the same level as that through the straight channels, even without tedious optimization of defect size and locations. For comparison, under the exact same coupling conditions, the measured transmission through a similar bending channel in the square-lattice (Fig. 4i) is found to be much narrower and lower, due to the mismatch of the propagating modes along the single row defect in the <100> direction, the single row defect in the <110> direction, and the horn antennas. Similar mismatch between the propagation modes along single-row defects in the <100> direction and the <001> direction is present in 3D woodpile photonic crystals (5).

Moreover, when a few roughly evenly spaced defect cylinders are placed inside the straight channel, a sharp resonant transmission peak, instead of a broad transmission band, appears. Importantly, the resonant frequency in these coupled resonant waveguides can be flexibly tuned by modifying the position of the defect cylinders. Two different sets of defect cylinders, marked as red or green dots in Fig. 4a, were used separately. Their corresponding transmission spectra are shown in Fig. 4c with red and green curves, respectively.  A rich variety of resonant cavity modes (for TM polarization) were found in simulation in a similar 2D hyperuniform disordered material made of dielectric rods and the Q factors were calculated to be  as high as $10^8$ (27). Thus, it seems likely that cavity-coupled resonator waveguides in the hyperuniform disordered structures may be finely tuned to act as a narrow band-pass filter with a high quality factor Q.

**Discussion**

This novel class of PBG material combines the advantages of isotropy due to disorder and controlled scattering properties due to low density fluctuations (hyperuniformity) and uniform local topology. The novel combination of these characteristics enables Mie resonances in individual cylinders to couple in "bonding" and "antibonding" modes that concentrate electrical field either in cylinders or in air cells separated by a band gap, reminiscent of the band edge states in periodic crystals. Our density of states simulation results (shown in Fig. 2) confirm that this is indeed a complete energy gap for photons (a forbidden frequency range) with the complete absence of states for any polarization, rather than a mobility gap associated with localized states or spatial band gap for certain wave-vectors along the direction perpendicular to the patterned plane (28). This photonic energy bandgap prohibits not only propagation, but also spontaneous emission of radiation at any gap frequency. Although in this paper we focus on 2D architectures, the same design principles can be applied to 3D (19).

In photonic crystals, efficient waveguides are limited to certain directions by crystal symmetries. Moreover, the mismatch between the propagation modes along line defects along different symmetry directions greatly limits the freedom of bending waveguides. Hence, until very recently there is only one successful 3D waveguiding demonstration in photonic crystals, which is proven to be strictly limited to bending from the <100> direction to the <101> direction (5). The design freedom associated with the intrinsic isotropy in our material is a significant advance over photonic crystal architectures.

In summary, we have used a novel constrained optimization method to engineer a new class of PBG materials, and have experimentally demonstrated for the first time two significant properties of these materials. We have proved the existence of an isotropic complete PBG (at all angles and for all polarizations) in an alumina-based 2D hyperuniform disordered material. Unlike photonic crystals, our material is disordered but still hyperuniform, lacking long-range translational order and Bragg scattering, yet resulting in an isotropic PBG. Furthermore, we have shown that the isotropic PBG enables the creation of freeform waveguides, impossible to obtain using photonic crystal architectures. These newly introduced waveguides can channel photons robustly in arbitrary directions with ready control of transmission bandwidth and can also be decorated with defects to produce sharply resonant structures useful for filtering and frequency splitting. These results demonstrate that hyperuniform disordered photonic materials may offer advantages to improve various technological applications which can benefit from a PBG (29,30) (e.g., displays, lasers (31), sensors (32), telecommunication devices (33), and optical micro-circuits (34)). Our findings are applicable to all wavelengths. Deep reactive ion etching on silicon or two-photon polymerization can be used to construct similar hyperuniform disordered structures with a PBG in the infrared or optical regimes. Our results also portend the creation of novel photonic, acoustic, and electronic materials with unprecedented physical properties unhindered by crystallinity and anisotropy.

**Materials and methods**
We have constructed the physical realization of a hyperuniform stealthy design using commercially available $Al_2O_3$ cylinders and walls cut to the designed heights and widths. The dielectric constant of these $Al_2O_3$ materials was measured to be 8.76 at the mid-gap frequency. The hyperuniform patterns consist of cylinders of radius $r$=2.5 mm connected by walls of thickness $t$=0.38 mm and with various widths to match the hyperuniform network; the components are 10.0 cm tall in the third dimension. The average inter-cylinder spacing is $a$=13.3 mm and the sample size used in our transmission measurements was $13a \times 13a$, corresponding to the region inside the red square shown in Fig 1a. A platform with the desired hyperuniform pattern with slots of depth 1 cm for the insertion of cylinders and walls was fabricated by stereolithography. A side view of the structure, Fig 1b, shows the patterned platform and the inserted cylinders and walls. Cylinders and walls can easily be removed and replaced to make cavities, waveguides, and resonance structures. Fig.1c is a photo of the structure viewed from above. Our experiments are carried out with microwaves in the spectral range of 7-13 GHz, $\lambda \sim 2a$, and with a setup similar to the one described in Ref. (35). The sample is placed between two facing microwave horn antennas. For bandgap measurements, the horns are set a distance of $28a$ apart to approximate plane waves. Absorbing materials are used around the samples to reduce noise.

Our theoretical band structure calculations were obtained using a supercell approximation and the conventional plane-wave expansion method (3,24). The size of the supercell used in the simulations is $22a \times 22a$ (the entire region of Fig.1a). We solve the vectorial Maxwell equations, assuming the structure is infinitely long in the vertical direction. The supercell's first Brillouin zone is then discretized in 64x64 k-points, and the band structure is evaluated on the k-space mesh. The calculated band structures for the TE and TM modes of our system are included in the supplementary materials. Bandgap boundaries are determined from these band structures and were confirmed to converge with several different realizations of hyperuniform disorder and larger supercell sizes up to $63a \times 63a$. We employ a Brillouin-zone integration scheme, similar to the one presented in Ref. (36), to evaluate the density of states (DOS).

**Acknowledgements:** This work was partially supported by the Research Corporation for Science Advancement (Grant 10626 to W. M.), the San Francisco State University start-up fund to W. M., the University of Surrey's support to M. F. (FRSF and Santander awards), and the National Science Foundation (NYU-MRSEC Program award DMR-0820341 to P.M.C, DMR-0606415 to ST, and ECCS-1041083 to P.S.J and M.F.). We thank Dr. Norman Jarosik for help and discussion on microwave measurements. We thank Mr. Daniel Cuneo for some computer support.

**Author Contributions:** W. M. directed the project, designed experiments, performed experiments, analyzed data and wrote the paper; M. F. initiated the project, performed numerical simulation and wrote the paper; E. W., Y. H., S. H., B. L., and D. L. set up the experiments, prepared the sample, performed the experiments, and analyzed data;. P. M. C. initiated the project, contributed to experimental design, and wrote the paper, S. T., and P. J. S initiated the project and wrote the paper.

**Figures:**

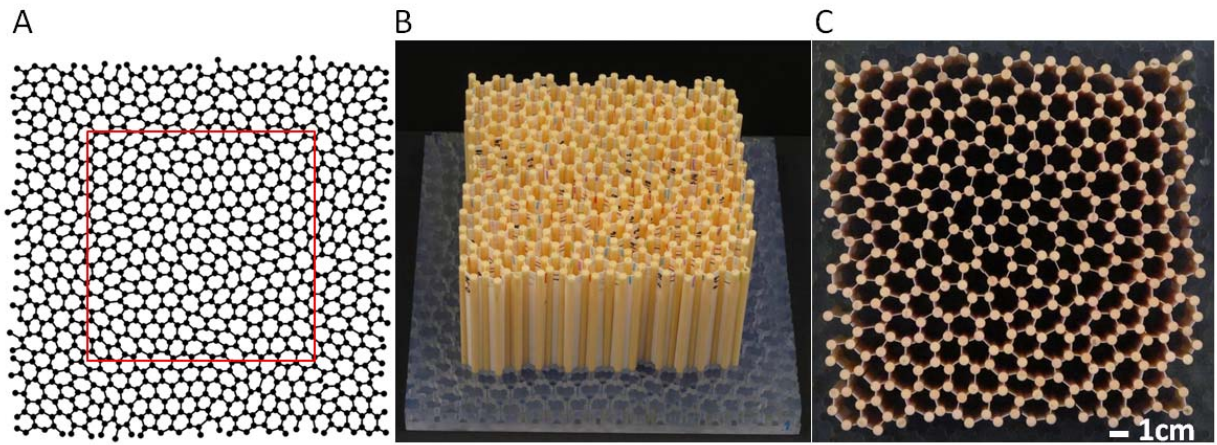

Figure 1. Design and photos of the hyperuniform disordered structure. (a) Cross-section of the 2D hyperuniform disorded structure, decorated with cylinders and walls. The area enclosed in the red box is the structure used for our experimental study. Side view (b) and top view (c) photographs of the hyperuniform disordered structure used in our experiment, assembled with $Al_2O_3$ cylinders and walls.

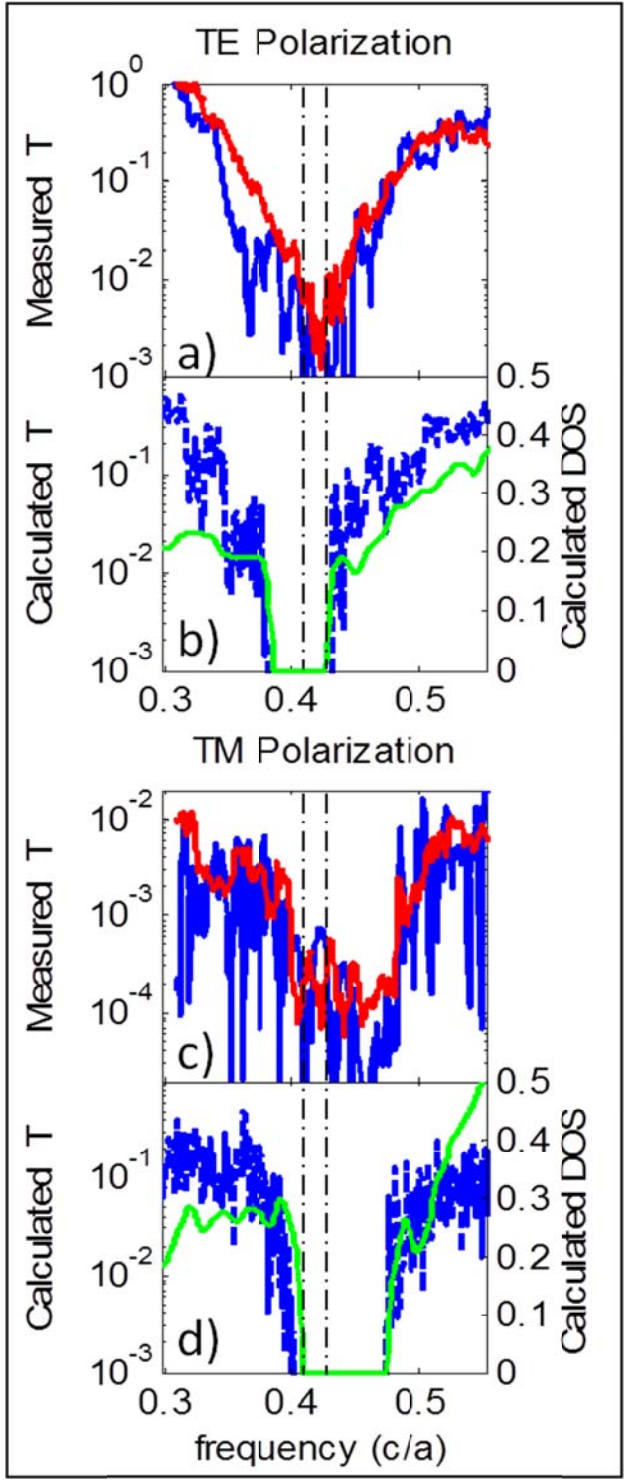

Figure 2. Measured and calculated transmission spectra and density of states (DOS) for the hyperuniform disordered sample. (a) Measured TE-polarization spectrum, at incident angle of zero degrees (blue) and averaged over all angles (red). (b) Calculated TE-polarization transmission spectrum (dashed blue) and calculated TE DOS (green). (c) Measured TM-polarization spectrum, at incident angle = 0 (blue) and averaged over all measured angles (red). (d) Calculated TM-polarization transmission spectrum (dashed blue) and calculated TM DOS (green). A -20 dB transmission drop from the measured maximum serves as an indicator of the bandgap for our hyperuniform samples. A similar relative 20dB drop indicates the bandgap in the TM mode for the square-lattice photonic crystal shown in Fig.4d. The vertical lines indicate the complete bandgap edges (both polarizations) from the DOS calculations. Frequencies are in units of c/a, where c is the speed of light.

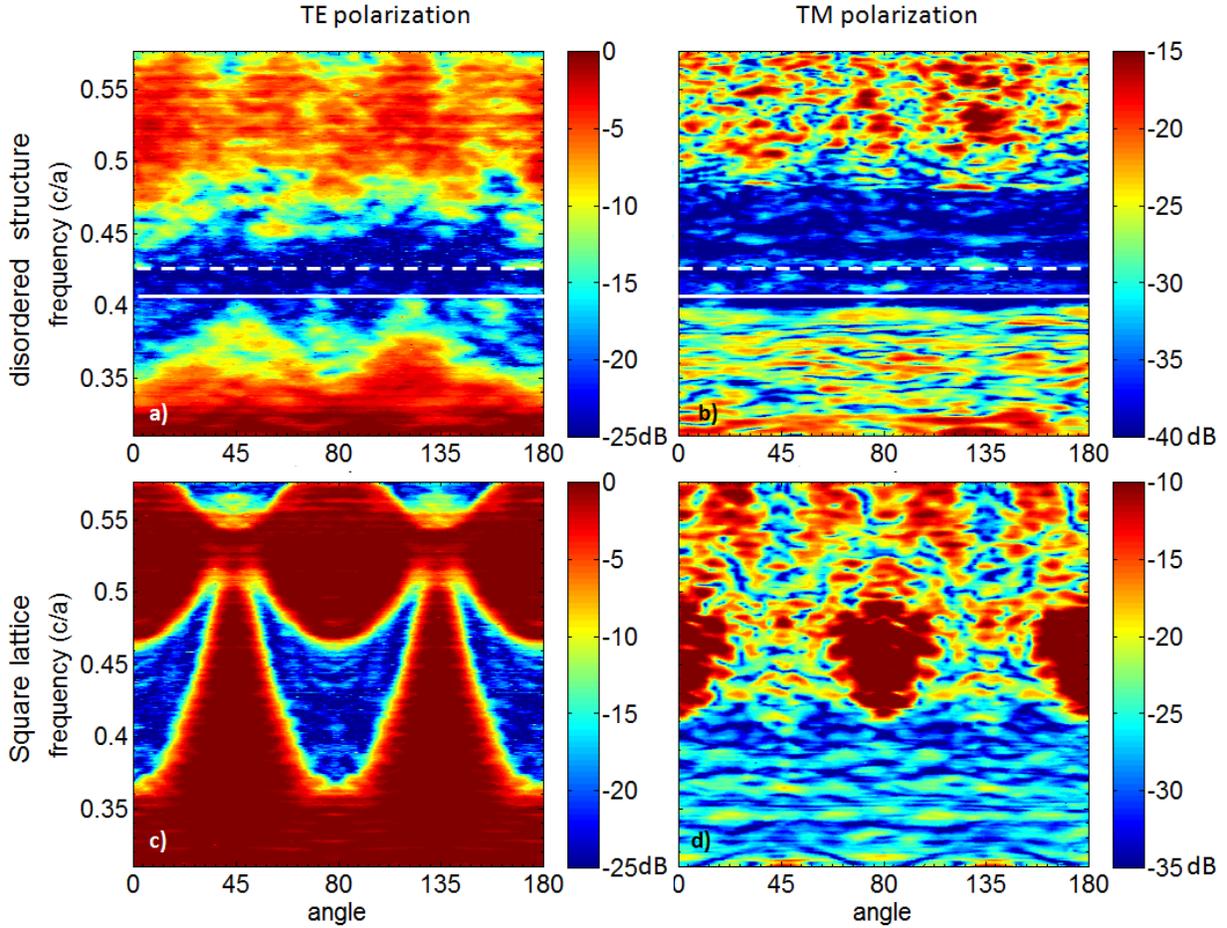

Figure 3. Measured transmission (color) as a function of frequency and incident angle. In the hyperuniform disordered structure, the measured bandgaps for TE (a) and TM (b) polarization overlap to form a complete PBG. The calculated boundaries of the complete PBG are shown with a solid white line (the lower boundary of the TM PBG) and a dashed white line (the upper boundary of the TE PBG). The measured transmission inside the calculated PBG drops by 20dB compared to the measured band pass maximum. In the square-lattice photonic crystal, stop gaps due to Bragg scattering occur along the Brillouin zone boundaries, varying dramatically with incident direction. For TE polarization (c), the stopgaps do not overlap in all directions so as to form a bandgap; for TM polarization (d), the stop gaps show an angular dependence associated with 4-fold rotational symmetry but overlap in all directions with a transmission reduction of 20 dB to form a band gap.

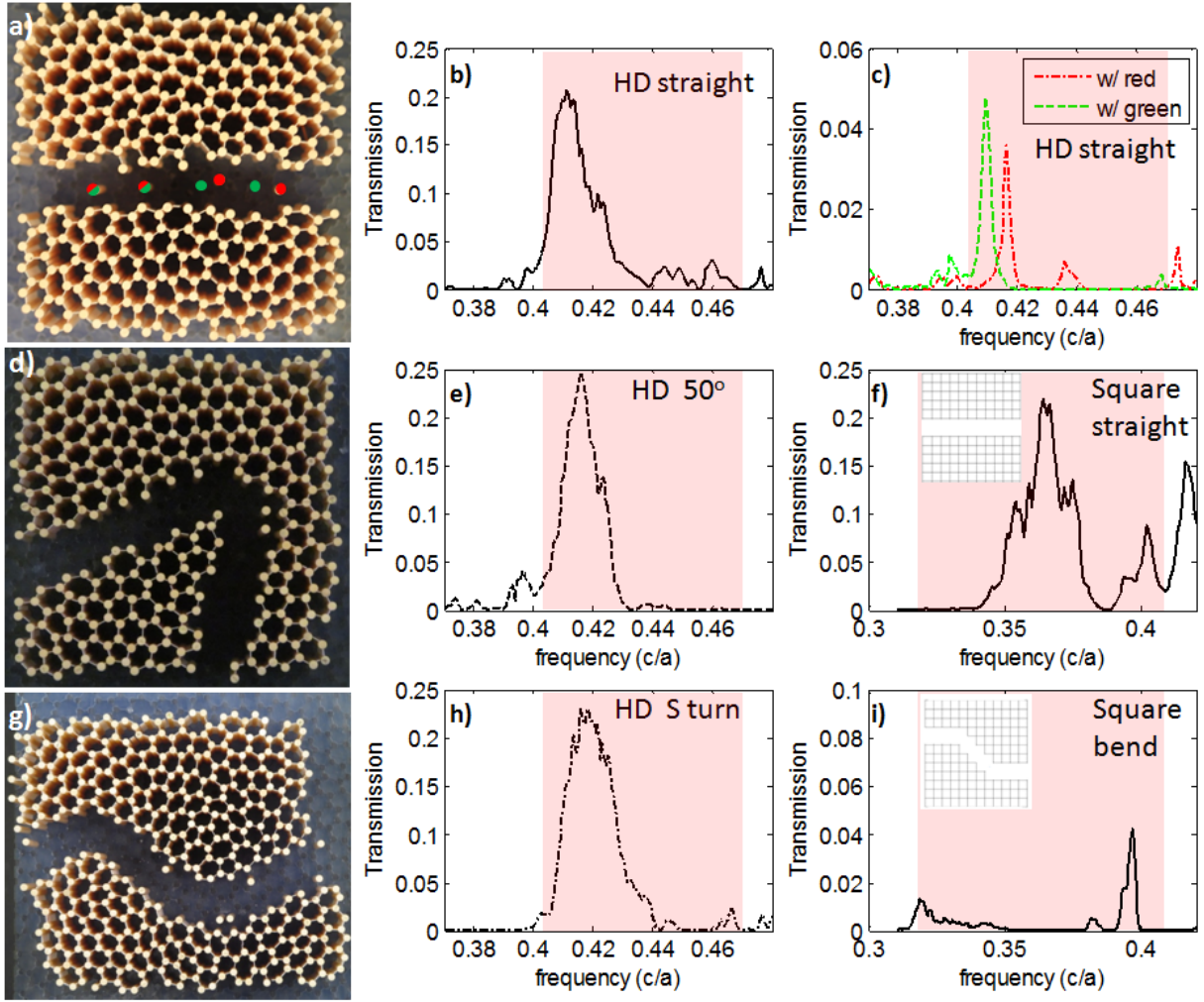

Figure 4. Measured TM-polarization transmission (detected power over source power) through different wave-guiding channels in the hyperuniform disordered structure and the square-lattice photonic crystal. Their respective TM-polarization bandgaps are highlighted with pink shading. a). Photograph of a straight channel of width $2a$ in the hyperuniform disordered structure. b). Measured TM transmission through the open straight channel in the hyperuniform disordered structure without extra defects. c). Measured TM transmission through the straight channel in the hyperuniform disordered structure, in which sets of four roughly evenly spaced defect cylinders are added to produce a narrow-band filtering channel. Two cases of defect locations (red or green dots in Fig. 4a) and their respective transmission (red or green curves) are shown. d). Photograph of a channel with a 50° bend. e). Measured TM transmission through the 50° bent channel. f). Measured transmission of a straight channel of width $2a$ (sketched in insert) in the square-lattice photonic crystal, which serves as a comparison to evaluate the performance of other channels. g). Photograph of a freeform "S" shaped channel. h). Measured TM transmission spectra through the "S" shaped channel. i). Measured transmission of a similar bending channel in the square-lattice photonic crystal (sketched in insert), created by removing one row of cylinders and their connected walls. The transmission is significantly lower and narrower than that through the bending channels in the hyperuniform structure, under the same coupling conditions.

**Supplementary figures:**

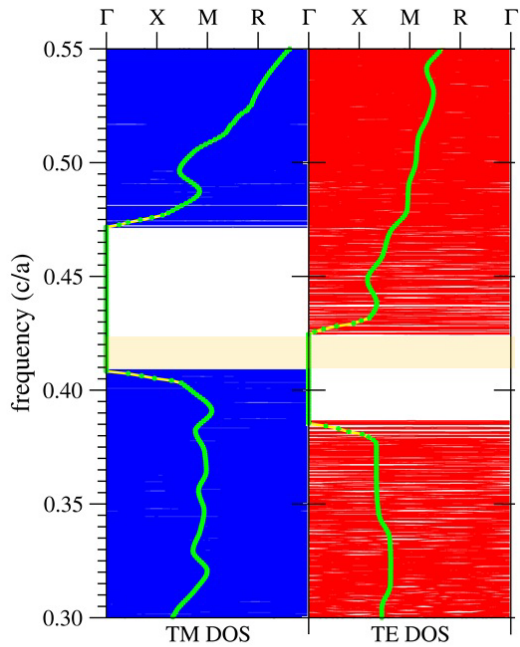

Figure S1. Simulations of TM (left) and TE (right) band structure (blue and red curves) and DOS (green curve) for the hyperuniform structure shown in Fig.1a. The gaps are identified as the region where the density of states is zero. The complete bandgap region is shown by the peach-colored area. The PBGs shown are equivalent to the fundamental band gap in periodic systems: For example, the spectral location of the TM gap in our structure is determined by the resonant frequencies of the scattering centers, and always occurs between band $N$ and $N + 1$, with $N$ precisely the number of cylinders per supercell. Similarly, for TE polarized radiation the band gap always occurs between bands $N$ and $N+1$, where $N$ is now the number of network cells in the structure.